\documentclass{conm-p-l}

\newtheorem{theorem}{Theorem}[section]

\theoremstyle{definition}
\newtheorem{definition}[theorem]{Definition}

\theoremstyle{remark}

\numberwithin{equation}{section}

\begin{document}
\title[Network of thin fibers]{Transition from a network of thin fibers to the quantum graph: an explicitly
solvable model}
\author{Stanislav Molchanov}
\address{Department of Mathematics and Statistics, University of North Carolina, %
\mbox{Charlotte}, North Carolina 28223}
\curraddr{Department of Mathematics and Statistics, University of North Carolina,
Charlotte, North Carolina 28223}
\thanks{The authors were supported in part by NSF Grant DMS-0405927.}
\author{Boris Vainberg}
\subjclass{}
\date{January 1, 1994 and, in revised form, June 22, 1994.}
\keywords{Quantum graph, wave guide, Dirichlet problem, asymptotics}

\begin{abstract}
We consider an explicitly solvable model (formulated in the Riemannian
geometry terms) for a stationary wave process in a specific thin domain $%
\Omega _{\varepsilon }$ with the Dirichlet boundary conditions on $\partial
\Omega _{\varepsilon }.$ The transition from the solutions of the scattering
problem on $\Omega _{\varepsilon }$ to the solutions of a problem on the
limiting quantum graph $\Gamma $ is studied. We calculate the Lagrangian
gluing conditions at vertices $v\in \Gamma $ for the problem on the limiting
graph. If the frequency of the incident wave is above the bottom of the
absolutely continuous spectrum, the gluing conditions are formulated in
terms of the scattering data of a problem in a neighborhood of each vertex $%
v\in \Gamma .$ Near the bottom of the absolutely continuous spectrum the
wave propagation is generically suppressed, and the gluing condition is
degenerate (any solution of the limiting problem is zero at each vertex).
\end{abstract}

\maketitle

2000]{Primary 35J05, 35P25, 58J37; Secondary 58J50}













\section{Introduction}

The paper concerns the asymptotic analysis of the wave propagation through a
system of wave guides (fibers) when the thickness $\varepsilon $ of the wave
guides is very small and the wave length is comparable to $\varepsilon $.
The simplest model one can consider is the stationary wave (Helmholtz)
equation
\begin{equation}
-\Delta u=\dfrac{\lambda }{\varepsilon ^{2}}u,\text{ \ \ \ }x\in \Omega
_{\varepsilon },  \label{h0}
\end{equation}
in the domain $\Omega _{\varepsilon }\subset R^{\nu },$ $\nu \geq 2,$
consisting of finitely many cylinders (tubes) of lengths $l_{1},l_{2},\cdots
l_{N}$ with the diameters of the cross-sections of order $O\left(
\varepsilon \right) $. Some of the lengths can be infinite. The junctions $%
J_{1},\cdots ,J_{N'}$ connecting the cylinders into networks are
compact
domains in $R^{\nu }$ with the diameters of the same order $O(\varepsilon )$%
. The axes of the cylinders and the centers of the junctions form edges and
vertices, respectively, of the limiting $\left( \varepsilon \rightarrow
0\right) $ metric graph $\Gamma $.

The Helmholtz equation in $\Omega _{\varepsilon }$ must be complemented by
the boundary conditions (BC) on $\partial \Omega _{\varepsilon }$. In some
cases (for instance, in a study of heat transport in $\Omega _{\varepsilon }$%
) the Neumann BC is natural. In fact, the Neumann BC presents the simplest
case due to the existence of a simple ground state (a constant) of the
problem in $\Omega _{\varepsilon }.$ \ However, in many applications, the
Dirichlet, Robin or impedance BC are more important. We will consider (apart
from a general discussion) only the Dirichlet BC, but all the arguments and
results can be modified to be applied to the problem with arbitrary BC.

We are going to study an explicitly solvable model. We assume that all the
tubes have the same cross sections $\omega _{\varepsilon }.\ $We also assume
that $\omega _{\varepsilon }$ is $\varepsilon -$homothety of a domain $%
\omega \in R^{\nu -1}.$ \ Let $\lambda _{0}$ be the principal eigenvalue of
the Laplacian $H_{0}=-\Delta _{\nu -1}$ in $\omega .$ Thus, $\varepsilon
^{-2}\lambda _{0}$ is the principal eigenvalue of $-\Delta _{\nu -1}$ in $%
\omega _{\varepsilon }.$ $\ $In the presence of infinite fibers, the
spectrum of the Dirichlet Laplacian on $\Omega _{\varepsilon }$ has an
absolutely continuous component which coincides with the semi-bounded
interval $[\varepsilon ^{-2}\lambda _{0},\infty )$. The equation (\ref{h0})
is considered under the assumption that $\lambda \geq \lambda _{0},$ when
propagation of waves is possible. There are two very different cases: $%
\lambda \rightarrow \lambda _{0}\ $as $\varepsilon \rightarrow 0,$ i.e. the
frequency is at the edge (or bottom) of the absolutely continuous spectrum,
or $\lambda \rightarrow \widehat{\lambda }>\lambda _{0},$ i.e. the frequency
is above the bottom of the absolutely continuous spectrum.

If $\varepsilon \rightarrow 0$, one can expect that the solution $%
u_{\varepsilon }$ of (\ref{h0}) on $\Omega _{\varepsilon }$ is close to the
solution $\widetilde{u}=\widetilde{u}_{\varepsilon }$ of a much simpler
problem on the graph $\Gamma $. The function $\widetilde{u}$ satisfies the
following equation on each edge of the graph
\begin{equation}
-\dfrac{d^{2}\widetilde{u}\left( s\right) }{ds^{2}}=\left( \dfrac{\widehat{%
\lambda }-\lambda _{0}}{\varepsilon ^{2}}\right) \widetilde{u},  \label{lsg}
\end{equation}
where $s$ is the length parameter on the edges. One has to add appropriate
gluing conditions (GC) on the vertices $v$ of $\Gamma $. These gluing
conditions give basic information on the propagation of waves through the
junctions. They define the solution $\widetilde{u}$ of the problem (\ref{lsg}%
) on the limiting graph. The ordinary differential equation (\ref{lsg}), the
GC, and the solution $\widetilde{u}$ depend on $\varepsilon .$ However, we
shall often call the corresponding problem on the graph the limiting
problem, since it enables one to find the main term of small $\varepsilon$
asymptotics for the solution $u=u_{\varepsilon }$ of the problem (\ref{h0})
in $\Omega _{\varepsilon }.$

The convergence of the spectrum of the problem in $\Omega _{\varepsilon }$
to the spectrum of a problem on the limiting graph has been extensively
discussed in physical and mathematical literature (e.g., \cite{DE}-\cite{EP}%
, \cite{K,KZ1,KZ2,P,RS} and references therein). This list, containing
important contributions to the topic and some review papers, is far from
complete. What makes our paper different is the following: all the
publications that we are aware of, are devoted to the convergence of the
spectra (or resolvents) only in a small (in fact, shrinking with $%
\varepsilon \rightarrow 0)$ neighborhood of $\lambda _{0}$ (bottom of the
absolutely continuous spectrum), or below $\lambda _{0}$. Usually, the
Neumann BC is assumed. We deal with asymptotic behavior of solutions of the
scattering problem in $\Omega _{\varepsilon }$ when $\lambda $ is close to $%
\widehat{\lambda }>\lambda _{0},$ and the BC on $\partial \Omega
_{\varepsilon }$ can be arbitrary. It turns out that the GC on the limiting
graph in the case $\lambda \rightarrow \widehat{\lambda }>\lambda _{0}$ is
different from the case when $\lambda \rightarrow \lambda _{0}.$

Papers \cite{FW}, \cite{KZ1}, \cite{KZ2}, \cite{RS} contain the gluing
conditions and the justification of the limiting procedure $\varepsilon
\rightarrow 0$ in the case when the Neumann BC is imposed at the boundary of
$\Omega _{\varepsilon },$ and $\lambda \rightarrow 0$ ($\lambda _{0}=0$ in
the case of the Neumann BC). Typically, the GC in this case are:$\;$the
continuity of $\widetilde{u}\left( s\right) $ at each vertex $v$ and $%
\sum_{j=1}^{d}\widetilde{u}_{j}^{\prime }\left( v\right) =0$, i.e.
the continuity of both the field and the flow. These GC are called
Kirchhoff's GC. In the case when the shrinkage rate of the volume
of the junction neighborhoods is lower than the one of the area of
the cross-sections of the guides,
more complex energy dependent or decoupling condition can arise (see \cite{K}%
, \cite{KZ2}, \cite{EP} for details).

Let us stress again that this is the situation near the bottom $\lambda
_{0}=0$ of the absolutely continuous spectrum. A recent paper by O. Post
\cite{P} contains analysis of the Dirichlet Laplacian near the bottom of the
absolutely continuous spectrum $\lambda _{0}>0$ under the condition that the
junction is more narrow than the tubes. It was proved in \cite{P} that in
this case the GC for the problem on the limiting graph are the Dirichlet
conditions, i.e. waves do not propagate through the narrow junction when $%
\lambda $ is close to the bottom of the absolutely continuous spectrum.

Our paper concerns the asymptotic analysis of the Dirichlet problem when $%
\lambda $ is close to $\widehat{\lambda }>\lambda _{0}.$ As we shall see,
the limiting GC conditions in this case differ from the Dirichlet or
Kirchhoff's conditions. They are formulated in terms of the scattering
coefficients of the original problem. At the first glance, it looks like a
vicious circle: one needs to solve the problem in $\Omega _{\varepsilon }$
and find the scattering coefficients in order to state the limiting problem.
However, this is not the case. The scattering coefficients are needed only
for local problems in domains which consist of one junction and adjoint
tubes, but not for the problem in $\Omega _{\varepsilon }.$

We emphasize again that we do not consider equation (\ref{h0}) in
a branching domain in Euclidean space, but rather a similar
equation on a product of a compact manifold and a quantum graph.
The latter equation admits a separation of variables. The proposed
model does not have direct physical significance. However, the
changes in the metric (or potentials) in our model play the role
of junctions, and the problem under consideration is a simplified
analog of a more realistic situation. The present model is chosen
only for the sake of simplicity. We proved that the obtained
results are valid for the problem in more general domains $\Omega
_{\varepsilon }$ (where the separation of variables is
impossible). This general case will be discussed elsewhere.

We also studied the Dirichlet problem for general domains $\Omega
_{\varepsilon }$ without special assumptions on the geometry of the
junctions when, simultaneously, $\varepsilon \rightarrow 0,$ $\lambda
\rightarrow \lambda _{0},$ and the diameters of the guides and junctions
have the same order $O(\varepsilon ).$ Our conclusion is that, generically,
the limiting GC in this case is the Dirichlet condition. Thus (generically)
waves will not propagate through the junctions when the frequency is close
to the bottom of the absolutely continuous spectrum. Let us stress that this
is true both in the case when the diameters of the junctions are smaller
than the diameters of the guides, and in the case when they are larger. Some
special conditions (they will be described in another paper) must be
satisfied for waves to propagate if $\lambda \rightarrow \lambda _{0}$. An
infinite cylinder, which can be considered as two half-infinite tubes with
the junction of the same shape, can be considered as an example of a domain
where the propagation of the wave at $\lambda =\lambda _{0}$ is not
suppressed. Practically, we do not deal with the problem near the bottom of
the absolutely continuous spectrum in this publication. A detailed analysis
of this problem will be published elsewhere. However, we show here that the
GC on the limiting graph for our simplified model with $\lambda =\widehat{%
\lambda }$, generically, have a limit as $\widehat{\lambda }\rightarrow
\lambda _{0}$ and the limiting conditions are the Dirichlet conditions.

The authors are grateful to P.~Exner, P.~Kuchment, and S.~Novikov for very
useful discussions.

\section{Setup of the model.}

To understand better the general model under consideration we start with a
simple example. Let us consider the Helmholtz equation
\begin{equation}
-\Delta u=\frac{\lambda }{\varepsilon ^{2}}u,\ u|_{\partial \Omega }=0
\label{or}
\end{equation}
inside of a thin $2$-dimensional strip $\Omega _{\varepsilon }$ which has a
constant width $\varepsilon $ outside of the $\varepsilon $-neighborhood of
the origin. This strip is a particular case of domains $\Omega _{\varepsilon
}$ described in the introduction. In our case $\Omega _{\varepsilon }$
consists of two half strips with one junction. Consider the case $%
\varepsilon =1.$ We roll up the strip $\Omega =\Omega _{1}$ and transform it
into a cylindrical surface $C$ with circle cross-sections. The radius of the
cross sections is constant outside of the neighborhood of the origin. Let $%
(\varphi ,s),$ $0\leq \varphi <2\pi ,$ $s\in R,$ be cylindrical coordinates
on $C,$ and let $A(s)$ be the radius of the cross-section of the cylinder $%
C. $ The boundary $\partial \Omega $ corresponds to a cut of the cylinder
along the meridian $\varphi =0$.

We replace the Euclidean metric on $C$ by the metric of the surface of
revolution: $dl^{2}=ds^{2}+A^{2}\left( s\right) d\varphi ^{2},$ and we
replace the Laplacian $\Delta $ on $\Omega $ by the Laplace-Beltrami
operator on $C$ which is defined by the metric $dl^{2}.$ Then we arrive at
\begin{equation}
-\frac{\partial ^{2}u}{\partial s^{2}}-\frac{A^{\prime }\left( s\right) }{%
A\left( s\right) }\frac{\partial u}{\partial s}-\frac{1}{A^{2}\left(
s\right) }\frac{\partial ^{2}u}{\partial \varphi ^{2}}=\lambda u,\text{ \ \
\ \ }u(0,s)=u(2\pi ,s)=0.  \label{c2}
\end{equation}
For arbitrary $\varepsilon >0,$ we assume that $\Omega _{\varepsilon }$ is $%
\varepsilon $-homothety of $\Omega ,$ i.e. $\Omega _{\varepsilon }$ is the
image of $\Omega _{\varepsilon }$ under the action of the map $(s,\varphi
)\rightarrow \varepsilon ^{-1}(s,\varphi ).$ Then the problem (\ref{c2})
takes the form
\begin{equation}
-\frac{\partial ^{2}u}{\partial s^{2}}-\frac{A^{\prime }\left( s/\varepsilon
\right) }{\varepsilon A\left( s/\varepsilon \right) }\frac{\partial u}{%
\partial s}-\frac{1}{A^{2}\left( s/\varepsilon \right) }\frac{\partial ^{2}u%
}{\partial \varphi ^{2}}=\frac{\lambda }{\varepsilon ^{2}}u,\text{ \ \ \ \ }%
u(0,s)=u(2\pi /\varepsilon ,s)=0.  \label{c22}
\end{equation}
One can expect that the propagation of waves governed by this modified
equation is very similar to the one governed by the original model (\ref{or}%
). These two models also have similar levels of difficulty.

Our general model is a direct generalization of (\ref{c2}). This model is
still very special and does not include most of branching domains $\Omega
_{\varepsilon }$ in Euclidean spaces. However, we believe that the model
captures some qualitative characteristics of wave propagation in thin
structures. The main advantage of the model is that it allows the separation
of variables.

Let $\Gamma $ be a connected quantum graph (see \cite{K}-\cite{KS}) with a
finite number of vertices $\left\{ v_{i}\right\} =V$ and edges $\left\{
e_{j}\right\} =E$. It is assumed that each edge $e_{j}$ has a
parameterization $s\in (0,l_{j}),$ $1\leq j\leq N,$ and some edges can have
infinite length. One can introduce the basic Hamiltonian $H_{0}=-\dfrac{d^{2}%
}{ds^{2}}$ on the space $C_{0}^{\infty }(\Gamma \backslash V)$ of infinitely
smooth functions, supported outside the set $V$. The space $L^{2}\left(
\Gamma \right) $ of square integrable functions on $\Gamma $ and the Sobolev
space $H^{1}\left( \Gamma \right) $ can be introduced in a natural way.
Later we will introduce self adjoint extensions of the operator $H_{0}$
which are defined by GC at vertices $v\in V.$

Let $M$ be a Riemannian $(\nu -1)$-dimensional manifold with the Riemannian
metric $d\varphi ^{2}$ and nontrivial boundary $\partial M$. Let $%
M_{\varepsilon }$ be the image of $M$ under the action of the map $\varphi
\rightarrow \varphi /\varepsilon .$ The equation will be given on the
manifold $\Omega _{\varepsilon }=\Gamma \times M_{\varepsilon }$ which is a
Cartesian product of a quantum graph $\Gamma $ and $M_{\varepsilon }.$ We
change the metric on $\Omega _{\varepsilon }$ in order to take into account
junctions between tubes $e_{j}\times M_{\varepsilon }.$ It is more
convenient to introduce first the metric on the rescaled manifold $\Omega
=\Gamma ^{\varepsilon }\times M$ where edges $e_{j}^{\varepsilon }$ of $%
\Gamma ^{\varepsilon }$ have lengths $l_{j}/\varepsilon .$ The Riemannian
metric $dl^{2}$ on $\Omega $ has the same form as in the example above, i.e.
for each edge $e_{j}^{\varepsilon }\in \Gamma ^{\varepsilon },$
\begin{equation}
dl^{2}=ds^{2}+A_{j}\left( s\right) ^{2}d\varphi ^{2},\text{ \ \ \ }s\in
(0,l_{j}/\varepsilon )  \label{dl}
\end{equation}
Here $A_{j}$ are smooth positive functions on $e_{j}^{\varepsilon }$ such
that $A_{j}(\xi )=1$ for $\xi \in (1,\frac{l_{j}}{\varepsilon }-1).$ The
Laplace--Beltrami operator on $\Omega $ associated with the metric $dl^{2}$
has the following form on each part $e_{j}^{\varepsilon }\times M$ of $%
\Omega :$
\begin{equation*}
\Delta =\frac{\partial ^{2}}{\partial s^{2}}+\frac{A_{j}^{\prime }\left(
s\right) }{A_{j}\left( s\right) }\frac{\partial }{\partial s}+\frac{1}{%
A_{j}^{2}\left( s\right) }\Delta _{\varphi },
\end{equation*}
where $\Delta _{\varphi }$ is the Laplace-Beltrami operator on $M.$ Thus,
the equation $-\Delta u=\lambda u$ on $\Omega ,$ after the inverse
rescaling, corresponds to the following equation on $\Omega _{\varepsilon }:$%
\begin{equation}
-\frac{\partial ^{2}u}{\partial s^{2}}-\frac{A_{j}^{\prime }\left(
s/\varepsilon \right) }{\varepsilon A_{j}\left( s/\varepsilon \right) }\frac{%
\partial u}{\partial s}-\frac{1}{A_{j}^{2}\left( s/\varepsilon \right) }%
\Delta _{\varphi }u=\frac{\lambda }{\varepsilon ^{2}}u,\text{ \ \ }e_{j}\in
E.  \label{ge}
\end{equation}
One could arrive to the equation (\ref{ge}) by introducing the appropriate
Riemannian metric directly on $\Omega _{\varepsilon }$ without rescalings
made above.

Equation (\ref{ge}) is symmetric with respect to the measure $d\sigma
=A_{j}(s/\varepsilon )dsd\varphi .$ The equation can be simplified (reduced
to an equation, which \ is symmetric with respect to the measure $dsd\varphi
$) using the substitution
\begin{equation*}
u(s,\varphi )\rightarrow A_{j}^{-1/2}(s/\varepsilon )u(s,\varphi ),\text{ \
\ \ \ }e_{j}\in E.
\end{equation*}
The new equation (we preserve the same notation $u$ for the solutions after
the substitution) has the form
\begin{equation}
-\frac{\partial ^{2}u}{\partial s^{2}}-\frac{1}{\varepsilon
^{2}A_{j}^{2}\left( s/\varepsilon \right) }\Delta _{\varphi }u+\varepsilon
^{-2}\left[ \frac{A_{j}^{\prime \prime }(s/\varepsilon )}{%
2A_{j}(s/\varepsilon )}-\frac{1}{4}(\frac{A_{j}^{\prime }(s/\varepsilon )}{%
A_{j}(s/\varepsilon )})^{2}\right] u=\frac{\lambda }{\varepsilon ^{2}}u,%
\text{ \ \ }e_{j}\in E.  \label{ge1}
\end{equation}

The equations (\ref{ge1}) have to be complemented by the appropriate BC, GC
and conditions at infinity. We impose the Dirichlet BC on $\Gamma
^{\varepsilon }\times \partial M:$%
\begin{equation}
u=0\text{ \ \ on \ }\Gamma ^{\varepsilon }\times \partial M.  \label{do}
\end{equation}
Let us recall that the degree $d=d(v)$ of a vertex $v$ is the number of
edges with an end point at $v$. We split the set $V$ of vertices in two
subsets $V=V_{1}\cup V_{2},$ where the vertices from the set $V_{1}$ have
degree $1$ and the vertices from the set $V_{2}$ have degree at least two.
We impose an arbitrary symmetric homogeneous BC on $v\times M,$ $v\in V_{1},$
for example the Dirichlet BC:
\begin{equation}
u=0\text{ \ \ on \ }v\times M,\text{ }v\in V_{1}.  \label{bbc}
\end{equation}

One has to specify the meaning of the equation (\ref{ge1}) on $v\times M$.
We assume that the solution $u$ of (\ref{ge}) satisfies Kirchhoff's GC\ at
vertices $v\in V_{2}$ , i.e. $u$ is continuous at each vertex $v\in V_{2}$
and the flow is preserved:
\begin{equation}
u\in C(\Omega _{\varepsilon }),\text{ \ \ \ }\sum_{j=1}^{d(v)}\frac{\partial
u}{\partial s_{j}}(v)=0,\text{ \ \ \ }v\in V_{2}.  \label{bcv}
\end{equation}
Here $\frac{\partial }{\partial s_{j}}$ is the differentiation with respect
to the parameter $s$ on the edge $e_{j}$ in the direction out of the vertex $%
v$. The conditions (\ref{bcv}) arise often in applications. The conditions (%
\ref{bcv}) are also in agreement with the GC in our first example, where $%
\Omega _{\varepsilon }$ is a strip and the functions on the strip are
smooth. Let us stress that Kirchhoff's GC\ at vertices $v\in V_{2}$ hold for
the solution of the problem in $\Omega _{\varepsilon }.$ As we will see, the
main term of asymptotics of that solution, which is determined by the
equations on $\Gamma ,$ satisfies different GC.

In order to complete the statement of the problem (\ref{ge1})-(\ref{bcv}),
one needs to specify conditions at infinity. We are going to consider
eigenfunctions and solutions of the scattering problems. To describe the
latter solutions, we note that the problem (\ref{ge})-(\ref{bcv}) admits the
separation of variables:
\begin{equation}
u(\gamma ,\varphi )=\psi (\gamma )\alpha (\varphi /\varepsilon ),  \label{rp}
\end{equation}
where $\gamma $ is a point on the graph $\Gamma $ and $\alpha (\varphi )$ is
an eigenfunction of the operator $\Delta _{\varphi }:$
\begin{equation*}
-\Delta _{\varphi }\alpha (\varphi )=\lambda ^{\prime }\alpha (\varphi ),%
\text{ \ \ }\varphi \in M;\text{\ \ \ \ }\alpha (\varphi )=0\text{ \ \ on \ }%
\partial M.
\end{equation*}
Obviously, $\alpha (\varphi /\varepsilon )$ is an eigenfunction of the
Dirichlet Laplacian on $M_{\varepsilon }$ with the eigenvalue $\lambda
^{\prime }/\varepsilon ^{2}.$ Then $\psi $ satisfies the equation
\begin{equation}
-\psi ^{\prime \prime }+\varepsilon ^{-2}Q(s/\varepsilon )\psi =(\frac{k}{%
\varepsilon })^{2}\psi ,\text{ \ \ \ \ \ \ }k=\sqrt{\lambda -\lambda
^{\prime }}>0,  \label{fe}
\end{equation}
where
\begin{equation*}
Q=Q_{j}=\frac{A_{j}^{\prime \prime }}{2A_{j}}-\frac{1}{4}(\frac{%
A_{j}^{\prime }}{A_{j}})^{2}+\lambda ^{\prime }(\frac{1-A_{j}^{2}}{A_{j}^{2}}%
)\text{ \ \ \ \ on }e_{j}\in E,
\end{equation*}
and the boundary conditions
\begin{eqnarray}
\psi &=&0\text{ \ \ at \ }v\in V_{1},  \label{bc2} \\
\psi &\in &C(\Gamma ),\text{ \ \ \ \ \ }\sum_{j=1}^{d}\frac{\partial \psi }{%
\partial s_{j}}(v)=0,\text{ \ \ \ }v\in V_{2}.  \label{bc1}
\end{eqnarray}

Let $E^{\prime }\subset E$ be the set of semi-bounded ($l_{j}=\infty $)
edges $\{e_{j}\},$ $1\leq j\leq r.$ We have a natural parameterization $s\in
(0,\infty )$ on $e_{j}\in E^{\prime }$ where $s=0$ corresponds to the end
point $v_{j}$ of $e_{j}$. Let us recall that the coefficients $A_{j}$ differ
from $1$ only in neighborhoods of the vertices, i.e. $A_{j}(s)=1,$ $s>1$ if\
$j\leq r.$ Then the equations (\ref{fe}) imply
\begin{equation*}
-\psi ^{\prime \prime }=\frac{k^{2}}{\varepsilon ^{2}}\psi \text{ \ \ on }%
e_{j}\in E^{\prime },\text{ \ \ }s>\varepsilon ;\text{ \ \ \ \ \ \ \ }k=%
\sqrt{\lambda -\lambda ^{\prime }}>0,
\end{equation*}
i. e. $\psi =\alpha e^{iks/\varepsilon }+\beta e^{-iks/\varepsilon }$ on $%
e_{j}\in E^{\prime },$ when $s>\varepsilon .$

\begin{definition}
The function $\psi ^{(m)}$, $1\leq m\leq r,$ is called a solution of the
scattering problem (on the graph) if it satisfies (\ref{fe})-(\ref{bc1}) and
\begin{equation}
\psi ^{(m)}=\left\{
\begin{array}{c}
e^{-iks/\varepsilon }+\tau _{m,m}e^{iks/\varepsilon }\text{ \ \ on }e_{m}\in
E^{\prime },\text{ \ \ }s>\varepsilon \\
\tau _{j,m}e^{iks/\varepsilon }\text{ \ \ on }e_{j}\in E^{\prime },\text{ \
\ }s>\varepsilon ,\text{ \ \ }j\neq m
\end{array}
\right. ,\text{ \ \ \ \ }k=\sqrt{\lambda -\lambda ^{\prime }}>0.  \label{ss}
\end{equation}
The function $u=u^{(m)}$ given by (\ref{rp}) with $\psi =\psi ^{(m)}$ is
called the solution of the scattering problem in $\Omega _{\varepsilon }.$
These functions describe the propagation of a wave of unit amplitude
incoming through the wave guide $e_{m}\times M,$ $\tau _{m,m}$ is the
reflection coefficient, $\tau _{j,m},$ $j\neq m,$ are the transmission
coefficients.
\end{definition}

Due to the separation of the variables, the scattering problem in $\Omega
_{\varepsilon }$ is reduced to the scattering problem on the graph. However,
this problem on the graph is governed by the equation (\ref{fe}) with
variable coefficients. Obviously, the potential $\varepsilon
^{-2}Q(s/\varepsilon )$ is supported in $\varepsilon $-neighborhoods of the
vertices.

Finally, we come to the main object of the investigation: asymptotic
analysis as $\varepsilon \rightarrow 0$ of the scattering solutions $%
u=u^{(m)}$. In fact, the next section contains asymptotic analysis of
arbitrary solutions of the problem (\ref{ge1})-(\ref{bbc}) for which
separation of variables (\ref{rp}) holds. For example, $u$ can be an
eigenfunction of the problem. Obviously, formula (\ref{rp}) reduces the
analysis of the asymptotic behavior of the function $u$ to the study of the
asymptotic behaviour of the solutions $\psi $ of the problem (\ref{fe})-(\ref
{bc1}).

\section{Asymptotic behavior of solutions.}

Let $\psi $ be an arbitrary solution of the problem (\ref{fe})-(\ref{bc1}).
The main feature of the Schr\"{o}dinger equation (\ref{fe}) is that the
potential is very singular (of order $\varepsilon ^{-2}$) near the vertices
and it vanishes outside of a very narrow (of the size $\varepsilon $)
neighborhood of the vertices. Then $\psi $ has the form $\psi =\alpha
_{j}e^{iks/\varepsilon }+\beta _{j}e^{-iks/\varepsilon }$ on each edge $%
e_{j}\in E$ outside of $\varepsilon $-neighborhoods of the vertices. We will
call the function $\widetilde{\psi }$ on $\Gamma ,$ which is equal to $%
\alpha _{j}e^{iks/\varepsilon }+\beta _{j}e^{-iks/\varepsilon }$ on each
edge $e_{j}\in E$ up to the end points, the ''limiting'' function. Let us
stress that the ''limiting'' function still depends on $\varepsilon ,$ and
this is the reason to use the word ''limiting'' in quotation marks. The
''limiting'' function provides a good simple approximation for the solution $%
\psi $ since it differs from $\psi $ only in $\varepsilon $-neighborhood of
the vertices. The goal of this section is to find a way to determine the
''limiting'' functions for scattering solutions and for eigenfunctions of
the problem on the graph directly (without solving the original problem).

The ''limiting'' function $\widetilde{\psi }$ for any solution $\psi $ of (%
\ref{fe})) satisfies the equation
\begin{equation*}
-\widetilde{\psi }^{\prime \prime }=(\frac{k}{\varepsilon })^{2}\widetilde{%
\psi }\text{ \ \ on }\Gamma .
\end{equation*}
However, if $\psi $ satisfies the GC (\ref{bc2}), the GC for $\widetilde{%
\psi }$ are different from (\ref{bc2}). In order to find the GC for $%
\widetilde{\psi }$ one needs to study supplementary problems for simple
subgraphs $\Gamma _{v}$ of $\Gamma $, consisting of one vertex $v$ of degree
$d=d(v)$ and the edges $e_{j_{1}},...,e_{j_{d}}$ with an end point at this
vertex. Let us consider one of such supplementary problems, when $d(v)>1$
(i.e. when $v\in V_{2}).$

We fix the parameterization on the edges $e_{j_{1}},...,e_{j_{d}}$ in such a
way that $s>0$ and $s=0$ corresponds to the point $v.$ Recall that the
potential $\varepsilon ^{-2}Q(s/\varepsilon ),$ defined on $\Gamma $,
vanishes outside $\varepsilon $-neighborhoods of the vertices. Let the
potential $\varepsilon ^{-2}B(s/\varepsilon )$ on $\Gamma _{v}$ coincide
with $\varepsilon ^{-2}Q(s/\varepsilon )$ in the $\varepsilon $-neighborhood
of the vertex $v$ and $B=0$ when $s>\varepsilon .$ Let $\varphi ^{(m)}$ be
solutions of the scattering problems for the Schr\"{o}dinger operator on the
subgraph $\Gamma _{v}$ with the potential $B:$%
\begin{equation}
-\varphi ^{\prime \prime }+\varepsilon ^{-2}B(s/\varepsilon )\varphi =(\frac{%
k}{\varepsilon })^{2}\varphi \text{ \ \ on }\Gamma _{v},  \label{b1}
\end{equation}
\begin{eqnarray}
\varphi &\in &C(\Gamma _{v}),\text{ \ \ \ \ \ }\sum_{j=1}^{d}\frac{\partial
\varphi _{n}}{\partial s}(v)=0,  \label{b2} \\
\varphi &=&\varphi ^{(m)}=\delta _{n,m}e^{-iks/\varepsilon
}+t_{n,m}e^{iks/\varepsilon }\text{ \ \ \ on }e_{j_{n}},\text{ \ \ }%
s>\varepsilon .  \label{b3}
\end{eqnarray}
Here $\varphi _{n}$ is the restriction of $\varphi $ on $e_{j_{n}},$ $\delta
_{n,m}$ is Kronecker's symbol. Of course, coefficients $t_{n,m}$ depend on $%
v.$ Let
\begin{equation*}
T_{v}=[t_{n,m}]
\end{equation*}
be the matrix of scattering coefficients. The diagonal elements $t_{m,m}$ of
the matrix $T_{v}$ are reflection coefficients of the corresponding
scattering solutions $\varphi ^{(m)},$ and the other elements are
transmission coefficients for $\varphi ^{(m)}$.

For any solution $\varphi $ of (\ref{b1}) and its ''limiting'' solution $%
\widetilde{\varphi },$ we denote by $\widehat{\varphi }$ the column vector
whose coordinates are restrictions $\widetilde{\varphi }_{n}$ of $\widetilde{%
\varphi }$ on $e_{j_{n}}.$

\begin{theorem}
\label{l3}Let $d(v)>1$ (i.e. $v\in V_{2}).$ Then

1) The matrix $T_{v}=[t_{n,m}]$ does not depend on $\varepsilon .$

2) For any function $\varphi $ which satisfies the equation (\ref{b1}) and
Kirchhoff's' GC (\ref{b2}), the ''limiting'' function $\widetilde{\varphi }$
satisfies the following GC at the vertex $v$:
\begin{equation}
\frac{i\varepsilon }{k}(I+T_{v})\frac{d\widehat{\varphi }}{ds}-(I-T_{v})%
\widehat{\varphi }=0,  \label{pgc}
\end{equation}
where $I$ is $d\times d$ unit matrix.

3) The matrix $T_{v}$ is unitary and symmetric ($t_{n,m}=t_{m,n}$); the $%
d\times 2d$ matrix $[\frac{\varepsilon }{k}(I+T_{v}),$ $(I-T_{v})]$ has rank
$d.$
\end{theorem}

\textbf{Remarks. }1) If the matrix $I+T_{v}$ is not degenerate, then (\ref
{pgc}) can be written in the form
\begin{equation*}
\frac{d\widehat{\varphi }}{ds}=C\widehat{\varphi },\text{ \ \ }C=\frac{-ik}{%
\varepsilon }(I+T_{v})^{-1}(I-T_{v}),
\end{equation*}
Here the matrix $C$ is Hermitian $(C=C^{\ast })$ due to the unitarity of $%
T_{v}$. If $I-T_{v}$ is not degenerate, then $\widehat{\varphi }=C^{-1}\frac{%
d\widehat{\varphi }}{ds}$ where $C^{-1}$ is Hermitian.

2) The scattering matrix $T_{v}$ can be expressed through the scattering
solutions of simpler scattering problems on individual edges, see below.

3) The unitarity of the scattering matrix is a standard fact in the
scattering theory. Its symmetry is also well known for 1-D Schr\"{o}dinger
equation. In 1-D case it means that the transmission coefficient does not
depend on the direction of the incident wave. The authors learned about the
symmetry of the scattering matrix in a more general situation from S.
Novikov (see \cite{N}).

\begin{proof}
In contrast to the case of the whole graph $\Gamma ,$ $\varepsilon $%
-independence of the scattering matrix $T_{v}$ for a star shaped graph $%
\Gamma _{v}$ becomes obvious after the rescaling $s\rightarrow s\varepsilon .
$

Let $\widetilde{\varphi }^{(m)}$ be the ''limiting'' function
which corresponds to the scattering solution $\varphi ^{(m)},$
i.e.
\begin{equation*}
\widetilde{\varphi }^{(m)}=\delta _{n,m}e^{-iks/\varepsilon
}+t_{n,m}e^{iks/\varepsilon }\text{ \ \ \ on }e_{j_{n}}\text{ \ for all\ }%
s\geq 0.
\end{equation*}
Then the Cauchy data for the corresponding vector $\widehat{\varphi }^{(m)}$
at the vertex $v$ are
\begin{equation*}
\widehat{\varphi }^{(m)}(v)=\delta _{m}+t_{m},\text{ \ \ \ }\frac{d\widehat{%
\varphi }^{(m)}}{ds}(v)=\frac{ik}{\varepsilon }(-\delta _{m}+t_{m}),
\end{equation*}
where $\delta _{m}$ is the column vector with coordinates $\delta _{n,m},$ $%
1\leq n\leq d,$ and $t_{m}$ is the $m$-th column of the matrix $T.$ Hence,
for the matrix $\Phi $ with the columns $\widehat{\varphi }^{(m)}(s),$ $%
1\leq m\leq d,$ we have
\begin{equation*}
\Phi (v)=I+T_{v},\text{ \ \ \ }\frac{d\Phi }{ds}(v)=\frac{ik}{\varepsilon }%
(-I+T_{v}).
\end{equation*}
This immediately implies that the ''limiting'' functions of\ the scattering
solutions $\varphi ^{(m)}$ satisfy the GC (\ref{pgc}).

It is obvious that the solution space for the problem (\ref{b1}), (\ref{b2})
is $d$-dimensional and that the scattering solutions are linearly
independent. Thus any solution $\varphi $ of (\ref{b1}), (\ref{b2}) is a
linear combination of the scattering solutions, and therefore its
''limiting'' function $\widetilde{\varphi }$ satisfies the GC (\ref{pgc}).

Let us prove the last statement of the theorem. If some of the
edges of the graph $\Gamma _{v}$ are finite, one can extend them
to infinity. Thus, without loss of the generality we can assume
that all these edges are infinite. Let $\Gamma _{v}^{(a)}$ be the
part of $\Gamma _{v}$ on which $s<a. $ Consider two scattering
solutions $\varphi ^{(m_{1})},$ $\varphi ^{(m_{2})} $ on $\Gamma
_{v}$ defined by (\ref{b1})-(\ref{b3}). Since the problem (\ref
{b1}), (\ref{b2}) is symmetric, from Green's formula for $\varphi
^{(m_{1})}$ and $\overline{\varphi }^{(m_{2})}$ on $\Gamma
_{v}^{(a)}$ it follows that
\begin{equation*}
\sum_{n=1}^{d}\left[ \frac{d\varphi _{n}^{(m_{1})}}{ds}\overline{\varphi }%
_{n}^{(m_{2})}-\varphi _{n}^{(m_{1})}\frac{d\overline{\varphi }_{n}^{(m_{2})}%
}{ds}\right] (a)=0.
\end{equation*}
If we substitute (\ref{b3}) in the last formula, we arrive at
\begin{equation*}
\sum_{n=1}^{d}t_{n,m_{1}}\overline{t}_{n,m_{2}}-\overline{t}%
_{n,m_{2}}e^{-2ika/\varepsilon }+t_{n,m_{1}}e^{2ika/\varepsilon }-\delta
_{m_{1},m_{2}}=0.
\end{equation*}
We take the average with respect to $a\in (A,2A)$ and pass to the limit as $%
A\rightarrow \infty .$ This leads to the orthogonality of\ the columns $%
t_{m_{1}},$ $t_{m_{2}}$\ of the matrix $T_{v}$ \ when $m_{1}\neq
m_{2}$ and to the condition $|t_{m_{1}}|=1$ when $m_{1}=m_{2}.$
Thus, the matrix $T_{v}$ is unitary. The symmetry of $T_{v}$ can
be proved similarly using Green formula for $\varphi ^{(m_{1})}$
and $\varphi ^{(m_{2})}$. Since the second part of the third
statement of the theorem is obvious, the proof of the theorem is
complete.
\end{proof}

Let now $v\in V_{1}$ and let $e_{v}$ be the edge of $\Gamma $ with an end
point at $v$. One can assume that the parameterization on $e_{v}$ is chosen
in such a way that $s=0$ corresponds to the vertex $v$.

We make the rescaling $s\rightarrow s\varepsilon $ on $e_{v}$ and consider
the following supplementary problem on the rescaled edge:
\begin{equation*}
-\varphi ^{\prime \prime }+Q(s)\varphi =k^{2}\varphi ,\text{ \ \ }s<\frac{%
l_{v}}{\varepsilon }-1;\text{ \ \ \ }\varphi (0)=0,\text{ \ }\varphi
^{\prime }(0)=1,
\end{equation*}
where $l_{v}=|e_{v}|$ is the length of $e_{v}.$ Then
\begin{equation*}
\varphi =\alpha _{v}e^{-iks}+\beta _{v}e^{iks},\text{ \ \ }s\in (1,\frac{%
l_{v}}{\varepsilon }-1),
\end{equation*}
where $\alpha _{v},$ $\beta _{v}$ do not depend on $\varepsilon .$

Let $\psi $ satisfy (\ref{fe}) on $e_{v}$ and $\psi (v)=0$. Then $\psi (s)$
on $e_{v},$ $s<l_{v}-\varepsilon ,$ is proportional to $\varphi
(s/\varepsilon ).$ Hence, the ''limiting'' function $\widetilde{\psi }$ has
the following form on $e_{v}$, $v\in V_{1}:$
\begin{equation*}
\widetilde{\psi }=C(\alpha _{v}e^{-iks/\varepsilon }+\beta
_{v}e^{iks/\varepsilon })\ \ \ \ \text{on }e_{v},\text{\ }%
s<l_{v}-\varepsilon ,
\end{equation*}
and therefore
\begin{equation}
\frac{i\varepsilon }{k}(\alpha _{v}+\beta _{v})\widetilde{\psi }^{\prime
}(v)-(\alpha _{v}-\beta _{v})\widetilde{\psi }(v)=0,\text{ \ \ }v\in V_{1}.
\label{fin}
\end{equation}

We note that Theorem \ref{l3} and the formula (\ref{fin}) provide a local
description of the GC for ''limiting'' functions, where the solution can be
defined only in a neighborhood of individual vertices. Thus, the following
theorem is an immediate consequence of the Theorem \ref{l3} and the formula (%
\ref{fin}).

\begin{theorem}
\label{t1}Let the function $\psi $ satisfy (\ref{fe}) and the Kirchhoff's'
GC (\ref{bc2}). Then the corresponding ''limiting'' function $\widetilde{%
\psi }$ is a solution of the equation
\begin{equation}
-\widetilde{\psi }^{\prime \prime }=(\frac{k}{\varepsilon })^{2}\widetilde{%
\psi }\text{ \ \ on }\Gamma ,  \label{111}
\end{equation}
which satisfies (\ref{fin}) and the GC
\begin{equation}
\frac{i\varepsilon }{k}(I+T_{v})\frac{d\widehat{\psi }}{ds}-(I-T_{v})%
\widehat{\psi }=0,\text{ \ \ \ }v\in V_{2}.  \label{222}
\end{equation}
In particular, if $\psi $ is a solution of the scattering problem (\ref{fe}%
)- (\ref{bc1}), then $\widetilde{\psi }$ is the solution of the scattering
problem for the equation (\ref{111}) with the conditions (\ref{fin}) and (%
\ref{222}). If $\psi $ is an eigenfunction of the problem (\ref{fe}), (\ref
{bc2}), then $\widetilde{\psi }$ \ is an eigenfunction of the problem (\ref
{111}), (\ref{fin}), (\ref{222}).
\end{theorem}

\section{Evaluation of the scattering matrix $T_{v}.$}

In the last part of the paper we will reduce the scattering problem (\ref{b1}%
)-(\ref{b3}) on the star shaped graph $\Gamma _{v}$ to $d$ simpler
scattering problems on individual edges $e_{j_{n}}$ extended to infinity in
both directions. Let the straight line $R$ with coordinates $s\in (-\infty
,\infty )$ represent the extended edges $e_{j_{n}}$. Let the potential $%
B_{n}(s)$ on $R$ coincide with the restriction of $B(s)$ on $e_{j_{n}}$ when
$s>0,$ and $B_{n}(s)=0$ for $s<0.$ Let $\psi =\psi _{(n)},$ $1\leq n\leq d,$
be the solution of the following scattering problem on the line $R$ (the
extension of the edge $e_{j_{n}}$):
\begin{eqnarray}
-\psi ^{\prime \prime }+B_{n}(s)\psi &=&k^{2}\psi ,\text{ \ \ \ \ }-\infty
<s<\infty ;  \notag \\
\psi &=&e^{iks}+r_{n}e^{-iks},\text{ \ \ }s<0;\text{ \ \ }\psi =t_{n}e^{iks},%
\text{ \ \ }s>1.  \label{qq}
\end{eqnarray}
This problem describes the propagation of the wave incoming from $s=-\infty $%
.

The following theorem allows one to express the scattering data $t_{n,m}$ of
the problem (\ref{b1})-(\ref{b3}) on the graph $\Gamma _{v}$ through the
reflection $r_{n}$ and transmission $t_{n}$ coefficients of the problem (\ref
{qq}). Let
\begin{equation*}
\rho =\sum_{j=1}^{d}\frac{1-r_{j}}{1+r_{j}}.
\end{equation*}

\begin{theorem}
\label{t2}The following formulas hold:
\begin{equation}
t_{n,m}=\frac{2t_{m}t_{n}}{(1+r_{m})(1+r_{n})\rho },\ m\neq n;\ \ \ \ \ \ \
t_{m,m}=\frac{2t_{m}^{2}}{(1+r_{m})^{2}\rho }-\frac{t_{m}(1+\overline{r}_{m})%
}{\overline{t}_{m}(1+r_{m})}.  \label{lf}
\end{equation}
\end{theorem}

\begin{proof}
Let us look for the solution $\varphi ^{(m)}$ (\ref{b1})-(\ref{b3}) on the
graph in the form
\begin{equation}
\varphi ^{(m)}(s)=\left\{
\begin{array}{c}
\gamma _{n}\psi _{n}(s/\varepsilon )\text{ \ \ \ \ on }e_{j_{n}},\text{ \ }%
n\neq m \\
\frac{1}{\overline{t}_{m}}\overline{\psi }_{m}(s/\varepsilon )+\gamma
_{m}\psi _{m}(s/\varepsilon )\text{\ \ \ \ on }e_{j_{m}}
\end{array}
\right.   \label{ass}
\end{equation}
with constants $\gamma _{n}$ which will be chosen below. In fact,
the constants $\gamma _{n}=\gamma _{n,m}$ depend also on $m,$ but
we will often omit the index $m$ to simplify formulas. Obviously,
$\varphi ^{(m)}$ satisfies (\ref{b1}) and (\ref{b3}) with
\begin{equation}
t_{n,m}=\gamma _{n,m}t_{n}.  \label{tmn}
\end{equation}
It remains only to choose $\gamma _{n}$ $=\gamma _{n,m}$in such a way that (%
\ref{b2}) holds. Then (\ref{ass}) will be the scattering solution of the
problem on the graph.

From (\ref{qq}) it follows that
\begin{equation*}
\psi _{n}(0)=1+r_{n},\text{ \ \ \ }\psi _{n}^{\prime }(0)=i(-1+r_{n}).
\end{equation*}
Thus, the substitution of (\ref{ass}) into (\ref{b2}) leads to the following
system
\begin{equation}
\gamma _{n}(1+r_{n})=\frac{1+\overline{r}_{m}}{\overline{t}_{m}}+\gamma
_{m}(1+r_{m}),\text{ \ }n\neq m;\text{ \ \ \ \ \ \ }\frac{1-\overline{r}_{m}%
}{\overline{t}_{m}}+\sum_{j=1}^{d}\gamma _{n}(-1+r_{n})=0.  \label{lee}
\end{equation}
Hence
\begin{equation}
\gamma _{n}=\gamma _{n,m}=\frac{1+\overline{r}_{m}}{\overline{t}_{m}(1+r_{n})%
}+\frac{\gamma _{m}(1+r_{m})}{(1+r_{n})},\text{ \ }n\neq m.  \label{gn}
\end{equation}
We substitute this expression for $\gamma _{n}$ into the last equation (\ref
{lee}) and arrive at the equation for $\gamma _{m}:$%
\begin{equation*}
\frac{1-\overline{r}_{m}}{\overline{t}_{m}}+\sum_{n\neq m}\frac{(1+\overline{%
r}_{m})(-1+r_{n})}{\overline{t}_{m}(1+r_{n})}-(1+r_{m})\rho \gamma _{m}=0,
\end{equation*}
which is equivalent to
\begin{equation}
\frac{1-\overline{r}_{m}}{\overline{t}_{m}}+\frac{(1+\overline{r}%
_{m})(1-r_{m})}{\overline{t}_{m}(1+r_{m})}-\frac{1+\overline{r}_{m}}{%
\overline{t}_{m}}\rho -(1+r_{m})\rho \gamma _{m}=0.  \label{zzz}
\end{equation}
The sum of the first two terms is equal to
\begin{equation*}
\frac{(1-\overline{r}_{m})(1+r_{m})+(1+\overline{r}_{m})(1-r_{m})}{\overline{%
t}_{m}(1+r_{m})}=2\frac{1-|r_{m}|^{2}}{\overline{t}_{m}(1+r_{m})}=\frac{%
2|t_{m}|^{2}}{\overline{t}_{m}(1+r_{m})}=\frac{2t_{m}}{(1+r_{m})}.
\end{equation*}
From here and (\ref{zzz}) it follows that.
\begin{equation*}
\gamma _{m}=\gamma _{m,m}=\frac{2t_{m}}{(1+r_{m})^{2}\rho }-\frac{1+%
\overline{r}_{m}}{\overline{t}_{m}(1+r_{m})}.
\end{equation*}
This together with (\ref{gn}) and (\ref{tmn}) implies (\ref{lf}). The proof
is complete.
\end{proof}

\section{Solutions near the bottom of the absolutely continuous spectrum.}

Recall that $\Gamma _{v}^{(a)}$ is the bounded part of the star shaped graph
$\Gamma _{v}$ on which $s<a.$ When $k=\sqrt{\lambda -\lambda ^{\prime }}>0,$
the GC\ at vertices $v\in \Gamma $ for the limiting problem on the graph are
formulated (see Theorem \ref{l3}) in terms of solutions of the scattering
problem (\ref{b1})-(\ref{b3}). In order to describe the behavior of the
solutions when $k=\sqrt{\lambda -\lambda ^{\prime }}\rightarrow 0,$ we need
to consider the Neumann problem on $\Gamma _{v}^{(a)}$ with the BC at points
where $s=a:$%
\begin{equation}
-\varphi ^{\prime \prime }+\varepsilon ^{-2}B(s/\varepsilon )\varphi =0\text{
\ \ on }\Gamma _{v}^{(a)},  \label{mm}
\end{equation}
\begin{equation}
\varphi \in C(\Gamma _{v}^{(a)}),\text{ \ \ \ \ \ }\sum_{j=1}^{d}\frac{%
\partial \varphi _{n}}{\partial s}(v)=0;\text{ \ \ }\frac{\partial \varphi
_{n}(a)}{\partial s}=0,\text{ \ }1\leq n\leq d.  \label{nn}
\end{equation}
The choice of $a$ above is not important since $B(s/\varepsilon )=0$ for $%
s>\varepsilon $ and solutions of (\ref{mm}), (\ref{nn}) are constants when $%
s>\varepsilon .$ For example, one can choose $a=2/\varepsilon .$ One can
also simplify the problem (\ref{mm}), (\ref{nn}) by changing the variable $%
s\rightarrow s\varepsilon .$ The function $\varphi $ in the new variable is
an eigenfunction with zero eigenvalue of the following Neumann problem:
\begin{equation}
-\varphi ^{\prime \prime }+B(s)\varphi =0\text{ \ \ on }\Gamma _{v}^{(2)},
\label{m1}
\end{equation}
\begin{equation}
\varphi \in C(\Gamma _{v}^{(a)}),\text{ \ \ \ \ \ }\sum_{j=1}^{d}\frac{%
\partial \varphi _{n}}{\partial s}(v)=0;\text{ \ \ }\frac{\partial \varphi
_{n}(2)}{\partial s}=0,\text{ \ }1\leq n\leq d.  \label{n1}
\end{equation}

Zero is not an eigenvalue of the problem (\ref{m1}), (\ref{n1}) for a
generic potential $B(s).$ Thus, the conclusion of the next theorem holds
generically.

\begin{theorem}
Let zero be not an eigenvalue of the problem (\ref{m1}), (\ref{n1}). Then
the GC (\ref{pgc}) ((\ref{222}), respectively) has the limit as $%
k\rightarrow 0,$ and the limit condition is the Dirichlet condition:\ $%
\varphi _{n}(v)=0,$ \ $1\leq n\leq d.$
\end{theorem}

\begin{proof}
Consider the restriction of the scattering solution $\varphi =\varphi ^{(m)}$
of the problem (\ref{b1})-(\ref{b3}) to $\Gamma _{v}^{(a)},$ $%
a=2/\varepsilon .$ After the rescaling $s\rightarrow s\varepsilon ,$ the
following relations hold
\begin{equation}
-\varphi ^{\prime \prime }+B(s)\varphi =k^{2}\varphi \text{ \ \ on }\Gamma
_{v}^{(2)},  \label{ff}
\end{equation}
\begin{equation}
\varphi \in C(\Gamma _{v}^{(a)}),\text{ \ \ \ \ \ }\sum_{j=1}^{d}\frac{%
\partial \varphi _{n}}{\partial s}(v)=0;\text{ \ \ }\frac{\partial \varphi
_{n}(2)}{\partial s}=-ik(\delta _{n,m}-t_{n,m}),\text{ \ }1\leq n\leq d.
\label{dd}
\end{equation}
Since $|t_{n,m}|\leq 1,$ from these relations it follows that $\varphi =O(k)$
on $\Gamma _{v}^{(2)}$ as $k\rightarrow 0.$ In particular, $\varphi
_{n}(2)=O(k)$ as $k\rightarrow 0.$ The latter together with (\ref{b3})
implies that
\begin{equation*}
\delta _{n,m}-t_{n,m}=O(k)\text{ \ \ as \ }k\rightarrow 0.
\end{equation*}
Differentiation of (\ref{ff}), (\ref{dd}) with respect to $k$ allows one to
conclude that $\frac{d\varphi _{n}}{dk}(2)=O(k)$ as $k\rightarrow 0.$ From
here it follows that $\frac{d}{dk}t_{n,m}=O(k)$ as $k\rightarrow 0.$ Hence, $%
I-T=O(k),$ $\frac{d}{dk}T=O(k),$ as $k\rightarrow 0.$  This allows one to
pass to the limit as $k\rightarrow 0$ in (\ref{pgc}), (\ref{222}) and get
the Dirichlet GC for the limiting function. The proof is complete.
\end{proof}

\end{document}